\documentclass[multicols,showpacs,preprintnumbers,amsmath,amssymb]{revtex4}

\newcommand{\SN}{S_{{\rm vN}}}
\newcommand{\BEQ}{\begin{eqnarray}}
\newcommand{\EEQ}{\end{eqnarray}}
\newcommand{\BEA}{\begin{eqnarray}}
\newcommand{\EEA}{\end{eqnarray}}

\renewcommand{\d}{{\rm d}}
\newcommand{\p}{\partial}

\newcommand{\half}{\frac{1}{2}}
\renewcommand{\thesection}{\arabic{section}}

\draft
\def\dbarrm {{\mathchar'26\mkern-11mu{\rm d}}}                       %
\begin{document} 
\draft
\title {Invalidity of the Landauer inequality for information erasure 
in the quantum regime}

\author{A.E. Allahverdyan$^{1,2)}$ and Th.M. Nieuwenhuizen$^{3)}$ } 
\address{
$^{1)}$ SPhT, CEA Saclay, 91191 Gif-sur-Yvette cedex, France\\
$^{2)}$ Yerevan Physics Institute, Alikhanian Brothers St. 2, 
Yerevan 375036, Armenia\\ 
$^{3)}$ Institute for Theoretical Physics, University of Amsterdam,
Valckenierstraat 65, 1018 XE Amsterdam, The Netherlands
}

\begin{abstract}
A known aspect of the Clausius inequality is that 
an equilibrium system subjected to a squeezing $\d S<0$ of its
entropy must release at least an amount $|\dbarrm Q|=T|\d S|$ of heat.
This serves as a basis for the Landauer principle, which puts 
a lower bound $T\ln 2$ for the heat generated by erasure of 
one bit of information. Here we show that in the world of quantum
entanglement this law is broken, suggesting that quantum 
carriers of information can be  more efficient than assumed so far.
\end{abstract}
\maketitle

\section{Introduction}


The laws of thermodynamics are at the basis of our
understanding of nature, so it is rather natural that 
they have applications beyond their original 
scope, e.g. in computing and information processing
\cite{rex,Neuman,leo,landauer,landauer1,ben,zurek,japan}. 
The first connection between information storage and thermodynamics 
was made by von Neumann in the 1950's~\cite{Neuman}. His speculation 
that each logical operation costs at least an amount of energy 
$T\ln 2$ was too pessimistic. Landauer pointed out that 
reversible ``one-to-one'' operations can be performed, in principle,
without dissipation; only irreversible operations ``many-to-one'' 
operations, like erasure, require dissipation of energy, 
at an amount at least equal to the von Neumann estimate 
$T\ln 2$ per erased bit~\cite{landauer,landauer1}. 
This conclusion is a direct consequence~\cite{japan} 
of the Clausius inequality, which connects the 
change of heat with the change of entropy.

The principal importance of erasure among other information-processing
operations arises because it is connected with changes of 
entropy, and thus cannot be realized in a closed system. One needs to couple
the information-carrying system with its environment. Therefore the process 
is accompanied with changes in heat, to be determined by thermodynamics. 
It was shown rigorously that all computations 
can be performed using reversible logical operations only~\cite{ben}.

Here we will consider thermodynamic aspects of erasure at low 
temperatures, so low that quantum effects start 
to play an important role. 
We choose the simplest example: a one-dimensional
Brownian particle in contact with a thermal bath at temperature $T$,
subject to an external confining potential.
The main new aspect arising at low temperatures is an
entanglement of the Brownian particle with the bath.
Therefore, even when the total system is in a pure state, the
subsystem (the Brownian particle) is in a mixed state.
Thus its stationary state cannot be given by equilibrium (Gibbsian)
quantum thermodynamics. We stress that
this situation is not at all exceptional, since it appears even
for a small but generic coupling provided that temperature is low enough.

Our main result will show that when entropy of 
the particle is decreased by external agents, 
namely when a part of the information carried by it is erased,
the particle can {\rm absorb heat} in clear contrast with the 
classical intuition.
Later we shall apply this result to show that there is not anything
similar to Landauer bound at low temperatures. 
Indeed, we point out that similar violation occurs for a spin
$\half$ particle coupled to a harmonic bath (spin-boson model).
Thus in this respect 
quantum carriers of information can be more efficient than
their classical analogs.

\renewcommand{\thesection}{\arabic{section}}
\section{Erasure of information and Gibbsian thermodynamics}
\setcounter{equation}{0}\setcounter{figure}{0} 
\renewcommand{\thesection}{\arabic{section}.}

Since information is carried by physical systems, messages are
coded by their states, namely every state (or possibly group
of states) corresponds to a ``letter''. The simplest example
is a two-state system, which carries on one bit of information.
The basic model of {\it source of information} in Shannonean, 
probabilistic information theory \cite{inf,inf1,jaynes} assumes that
the carrier of information can be in different states with certain
(so called a priori) probabilities. In other words, the messages of this
source appear randomly and the measure of their expectation is given 
by the corresponding probabilities. 
In the quantum case this situation is described by 
a density matrix $\rho $,
\BEA
&& \rho =\sum_n p_n |n\rangle \langle n|,\\
\label{tiger}
\EEA
which means that the carrier occupies a state $|n\rangle$ with the
a priori probability $p_n$. Moreover, different quantum states are 
exclusive, $\langle n|m\rangle = \delta _{nm}$.

The fundamental theorem by Shannon \cite{inf,inf1,jaynes} 
states that the information carried by an information source 
is given by its von Neumann entropy,
\BEA
\label{q2}
\SN (\rho )=-\sum_{n}p_n\ln p_n = -{\rm tr}(\rho \ln \rho ),
\EEA
The physical meaning of this result 
can be understood as follows. A source which has lower entropy
occupies fewer states with higher probability. It can be said 
to be better known, and therefore the appearance of its results
will bring less information. In contrast, a source with higher 
entropy occupies more states with lower probability. Its messages
are less expectable, and therefore bring more information.
The rigorous realization of this intuitive arguments appeared to 
be the most straightforward and fruitful proof of the Shannon 
theorem \cite{jaynes,balian}.
Notice that the entropy appears here on 
the information theoretical footing and not as purely 
thermodynamical quantities \cite{jaynes}.

Erasure is an operation which is done by an external agent in order
to reduce the entropy of the information carrier. This means 
that in its final state the carrier brings less information, i.e. 
some amount of it has been erased. 
In particular, a complete erasure corresponds to the minimization 
of entropy. 
Following standard
assumptions \cite{klim,balian} we will model external operations by a
time-dependent Hamiltonian $H(t)$ of the carrier, namely some of its 
parameters will be varied with time according to given trajectories.
If the information carrying system is closed, then its dynamics is
described by  the von Neumann equation 
\BEA
\label{kro}
\frac{\d}{\d t}\rho =\frac{i}{\hbar}[\,\rho (t)H(t)-H(t)\rho (t)\,]
\EEA
The entropy remains constant in time. 
In order to 
change it, one has to consider an information carrier, 
which is an open system. In that
case a part of its energy will be controlled (i.e. 
transferred or received) by its environment as {\it heat}. 
Indeed, the average energy of the carrier $U={\rm tr}\,H(t)\rho (t)$ 
 changes during a time $\d t$ as:
\begin{equation}
\label{dE}
\d U=\dbarrm {\cal Q}+\dbarrm {\cal W}
    = {\rm tr}[\,H\d \rho \,] +{\rm tr}[\,\rho\, \d H \,]
\end{equation}
The last term represents the averaged mechanical work
$\dbarrm {\cal W}$ produced by an external agent \cite{klim,balian}.
The first term in r.h.s. of Eq.~(\ref{dE}) arises
due to the statistical redistribution in 
phase space. We shall identify it with the change of heat 
$\dbarrm {\cal Q}$ \cite{klim,balian}, 
so Eq. (\ref{dE}) is just the first law of thermodynamics.

\renewcommand{\thesection}{\arabic{section}}
\section{Quantum Brownian particle 
in contact with a bath}
\setcounter{equation}{0}\setcounter{figure}{0} 
\renewcommand{\thesection}{\arabic{section}.}

As explained, at low temperatures of the bath the Brownian particle is not 
described by the quantum Gibbs distribution, except for very weak 
interaction with the bath. Therefore, its state
at low temperatures has to be found from first principles, starting 
from the microscopic description of the bath and the particle. This program
was realized in \cite{AN,AN1,weiss,haake}. In particular, in 
\cite{AN,AN1} we investigated statistical thermodynamics of the quantum 
Brownian particle.

Here we consider a simple example, a harmonic oscillator
with Hamiltonian 
\BEA
\label{ham}
H(p,x)=\frac{p^2}{2m}+\frac{ax^2}{2},
\EEA
where $m$ is the mass, and $a$ is the width.
The state of this particle can be described through the Wigner 
function \cite{balian}. In quantum theory this 
object plays nearly the same role as the common distribution 
of coordinate and momentum in the classical theory.
The stationary Wigner function reads \cite{grabert,weiss}\cite{AN,AN1}: 
\begin{eqnarray}
\label{ole77}
W(p,x)
=\frac{1}{2\pi}\sqrt{\frac{a}{mT_pT_x}}\,\,
\exp{[-\frac{p^2}{2mT_p}-\frac{ax^2}{2T_x}]},
\end{eqnarray}
where $T_x=a\langle x^2\rangle$ and $T_p={\langle p^2\rangle}/{m}$
are two effective temperatures, to be discussed a bit later.
Eq.~(\ref{ole77}) represents the state
of the particle, provided that the interaction with the bath was 
switched on long time before, so that the particle already came to 
its stationary state. 
The effective temperatures $T_p$ and 
$T_x$ depend not only on the system parameters $m$, $T$, and $a$,
but also on the damping constant $\gamma$, which quantifies
the interaction with the thermal bath, and on a large 
parameter $\Gamma$ which is the maximal characteristic
frequency of the bath. In particular, the Gibbsian limit
corresponds to $\gamma \to 0$. Then the distribution
(\ref{ole77}) tends to the quantum Gibbsian:
$T_x=T_p=\half\hbar\omega_0{\rm coth}(\half\beta\hbar\omega _0)$,
where $\omega _0=\sqrt{a/m}$.
In the classical limit, which is realized for $\hbar\to 0$ or $T\to
\infty$, the dependence on $\gamma $ and $\Gamma $ disappears; 
both $T_p$ and $T_x$ go to $T$.
The appearance of the effective temperatures 
in the quantum regime can be understood as follows.
For $T\to 0$ quantum Gibbs distribution predicts the pure vacuum state
for the particle. Due to quantum entanglement this cannot be the case
for a non-weakly interacting particle, so must $T_x$, $T_p$ depend on 
$\gamma $, and, being non-trivial, they have to be obtained from
first principles, as the state is not Gibbsian.
We will be interested by 
the so-called quasi-Ohmic limit where $\Gamma$ is the
largest characteristic frequency of the problem,
 the most realistic situation for 
information storing devices. In this limit one approximately has:
\begin{eqnarray}
\label{Tp}
&&T_p=\frac{\hbar}{\pi(\omega_1-\omega_2)}\left [
(\omega_1^2-\omega_2^2)\psi(\frac{\beta\hbar\Gamma}{2\pi})
-\omega_1^2\psi_1 
+\omega_2^2\psi_2\right ]-T\\
&&T_x=\frac{\hbar a}{m\pi(\omega _1-\omega _2)}
\left [\psi_1-\psi_2\right ]-T,
\qquad \psi_{1,2}\equiv\psi(\frac{\beta \hbar\omega _{1,2}}{2\pi})
\label{Tx}
\end{eqnarray}
where $\psi(z)=\Gamma'(z)/\Gamma(z)$ and
 $\omega _{1,2}=({\gamma}/{2m})(1\pm \sqrt{1-4am/\gamma^2 })$.

The average energy of the Brownian particle
\BEA
\label{w7}
U=\int \d x\d p W(p,x)H(p,x)=\frac{T_p}{2}+\frac{T_x}{2}
\EEA
depends on $a$ and $m$.
We will need the entropies 
\begin{eqnarray}
\label{S}
&&S_p= -\int \d p W(p)\ln W(p)=\half\ln (mT_p), \\
&&S_x=-\int \d x W(x)\ln W(x)=\half\ln \frac{T_x}{a}.
\label{S1}
\end{eqnarray}

The expressions for heat and work are generalized from Eqs.~(\ref{dE})
by simply using the Wigner function $W(p,x)$ instead of $\rho$.
One can prove by a direct calculation that quantities $T_p$, $T_x$
do deserve their nomenclature, since the classical Clausius equality can 
be generalized as
\begin{eqnarray}\label{asala2}
\d U=\dbarrm {\cal Q}+\dbarrm {\cal W}=
T_p\d S_p+T_x\d S_x+\dbarrm {\cal W},
\end{eqnarray}
for variation of any parameter. 
We will be especially interested in variation of the mass and the width
of the potential. The corresponding changes of heat read
\BEA
\label{khosrov}
&&\dbarrm{\cal Q}=
\half\left(\frac{\p T_p}{\p a}+\frac{\p T_x}{\p a}-\frac{T_x}{a}\right)\d a
+
\half\left(\frac{\p T_p}{\p m} + \frac{\p T_x}{\p m}+\frac{T_p}{m}\right)\d m.
\EEA
One can show that ${\p{\cal Q}}/{\p a}\le 0$, ${\p{\cal Q}}/{\p m}\ge 0$,
for all values of parameters including, of course, the classical limit.
The work done on the system is
\BEA
\label{khosrov2}
\frac{\p {\cal W}}{\p a}=\langle \frac{\partial H}{\partial a}\rangle
=\half\langle x^2\rangle\ge 0,\qquad
\frac{\p {\cal W}}{\p m}=\langle \frac{\partial H}{\partial m}\rangle
=-\half\frac{\langle p^2\rangle}{m^2}\le 0.
\EEA

The result for the von Neumann entropy (\ref{q2})
reads~\cite{weiss}
\BEA\label{SvN=}
S_{vN} =(w+\half)\ln(w+\half) -(w-\half)\ln(w-\half),
\quad w=\sqrt{\frac{mT_pT_x}{\hbar^2a}}.
\EEA

\renewcommand{\thesection}{\arabic{section}}
\section{Heat absorption with(out) entropy decrease}
\setcounter{equation}{0}\setcounter{figure}{0} 
\renewcommand{\thesection}{\arabic{section}.}

Now we will show that there are erasure processes, namely processes where
$\d \SN\le 0$, which are accompanied by an absorption of heat.
We noticed already that heat is always absorbed, when the mass is increased. 
There is a mass-increasing process, where $\d \SN\le 0$, since
one has at very low temperatures:
${\partial S_{vN} }/{\partial m}\sim{\partial w }/{\partial m}\le 0$.
An analogous argument can be brought about in the weak-coupling 
case, $\gamma\to 0$, and low-temperature limit:
\BEA
\label{ahmad}
T_p=\frac{\hbar\omega _0}{2}+\frac{\hbar\gamma}{\pi m} 
\ln\frac{\Gamma }{\omega _0\sqrt{e}},\qquad
T_x=\frac{\hbar\omega _0}{2}-\frac{\hbar\gamma}{2\pi m}. 
\EEA
This also implies
${\partial w }/{\partial m}<0$.
Recall that the corresponding expression
for $\p {\cal Q}/\p m$ was 
positive. This just means that for the variation of $m$ we have an 
interesting case where heat is absorbed when entropy is decreasing.
This is a counterexample
for the general validity of the Landauer principle.

In another approach we considered the spin-boson model~\cite{NMRprl},
where a spin-$\half$ particle is coupled to a bosonic bath.
There it is possible to give a pulse, where the spin
is rotated very fast, but, since it is only rotated,
 entropy is conserved.
Nevertheless, there are cases where heat can be 
extracted from the bath.

\renewcommand{\thesection}{\arabic{section}}
\section{Conclusion}
\setcounter{equation}{0}\setcounter{figure}{0} 
\renewcommand{\thesection}{\arabic{section}.}

The Landauer principle requires dissipation (release) of $T|\d S|$ units
of energy as a consequence of erasure of $|\d S|$ units of information.
This was believed to be the only {\it fundamental} energy cost of
computational processes \cite{landauer,landauer1,ben,japan}. 
Though in practice computers dissipate much more
energy, the Landauer principle was considered to put a general physical
bound to which every computational device interacting with its thermal 
environment must satisfy. Indeed, in several physical situations the Landauer
principle can be proved explicitly \cite{japan}. 

The main purpose of the present paper was to provide counterexamples of 
this principle, and thus to question its universal validity. In the reported 
case all general requirements on the information carrier and its interaction 
with the bath are met. The only new point of our approach is that we were
interested by sufficiently low temperatures, where quantum effects 
are relevant.
The Landauer principle appeared to be violated by these effects (in particular,
by entanglement). The result occurs in the two most standard models,
namely the Caldeira-Leggett model (central oscillator coupled to 
a harmonic bath) and the spin-boson model 
(two-level system coupled to such a bath), 
and therefore should be very general.

Recently the Landauer bound attracted serious attention 
from the field of applied information science \cite{ralf}. 
There is a belief that it
can be approached by further miniaturization of computational devices.
It is hoped that the present paper will help to understand limitations of
the Landauer principle itself, which may lead to unexpected
mechanisms for computing in the quantum regime.



\begin{thebibliography}{4}
	
\bibitem{rex} {\it Maxwell's demon: 
Entropy, Information, Computing }, ed. by H.S. Leff,
and A.F. Rex, Adam Hilger, 1990.

\bibitem{Neuman}J. von Neumann, {\it Theory of Self-reproducing
Automata}, Lect. 3 (Univ. Illinois Press, Urbana, IL, 1966).

\bibitem{leo}L. Szilard, Z. Phys. {\bf 53} 840 (1929);
reprinted in \cite{rex}.


\bibitem{landauer} R. Landauer, IBM J. Res. Dev. {\bf 5} 183 (1961);
reprinted in \cite{rex}.

\bibitem{landauer1}R. Landauer, Nature, {\bf 355} 779 (1988)

\bibitem{ben} C.H. Bennett, Int. J. Theor. Phys. 905 {\bf 21}, (1982);
reprinted in \cite{rex}.

\bibitem{zurek}W.H. Zurek, Nature {\bf 347} 119 (1989)

\bibitem{japan} K. Shizume, Phys. Rev. E {\bf 52} 3495 (1995)

\bibitem{AN} A.E. Allahverdyan and Th.M. Nieuwenhuizen, 
Phys. Rev. Lett {\bf 85}, 1799 (2000)

\bibitem{AN1} A.E. Allahverdyan and Th.M. Nieuwenhuizen, 
cond-mat/0011389

\bibitem{N1} Th.M. Nieuwenhuizen, 
Phys. Rev. Lett. {\bf 80} 5580 (1998) and
Phys. Rev. E {\bf 61} 267 (2000)

\bibitem{AN0} A.E. Allahverdyan and Th.M. Nieuwenhuizen, 
Phys. Rev. E {\bf 62} 845 (2000)

\bibitem{inf} 
I. Chisar and Y. Kerner, {\it Information Theory}, Academic Press, 1983.

\bibitem{inf1} Cover, T.N. \& Thomas, J.A. 
{\it Elements of Information Theory}, Academic Press, 1983.

\bibitem{jaynes} E. Jaynes, Phys. Rev. {\bf 108} 171 (1957);
{\it ibid} {\bf 108} 620 (1957); Am. J. Phys. {\bf 31} 66 (1963)


\bibitem{klim} 
J. Keizer, {\it Statistical Thermodynamics of Nonequilibrium
Processes}, (Springer-Verlag, 1987); 
Yu. L. Klimontovich,
{\it Statistical Theory of Open Systems}, (Kluwer, Amsterdam, 1997)


\bibitem{balian} R. Balian, {\it From Microphysics to Macrophysics}, 
I, II, (Springer-Verlag, 1992)


\bibitem{ford} 
G.W. Ford, M. Kac and P. Mazur, J. Math. Phys.,
{\bf 6} 504 (1965).

\bibitem{ullersma}
P. Ullersma, Physica {\bf 32} 27 (1966); ibid 56; 74; 90.

\bibitem{caldeira} A. Caldeira and A. Leggett, Phys. Rev. A {\bf 31}
1059 (1985).



\bibitem{weiss} U. Weiss, {\it Quantum Dissipative Systems}, 
(World  Scientific, Singapore, 1993).


\bibitem{haake}
F. Haake and R. Reibold, Phys. Rev. A {\bf 32} 2462 (1985).

\bibitem{grabert} H. Grabert, U. Weiss and P. Talkner,
Z. Phys. B {\bf 55} 87 (1984)



\bibitem{berger}  J. Berger, Int. J. Theor. Phys., {\bf 29} 985 (1990)

\bibitem{wolpert} D. Wolpert, Phys. Today, 98 {\bf 45} (1992)

\bibitem{NMRprl} A.E. Allahverdyan and Th.M. Nieuwenhuizen,
{\it Bath generated work extraction and inversion-free gain
 in two-level systems}, cond-mat/0201408


\bibitem{ralf}R.C. Merkle, Nanotechnology, {\bf 4} 21 (1993);
                                            {\bf 4} 114  (1993). 
\end{thebibliography}
\end{document}